\documentclass[aps, prl, twocolumn,showpacs,superscriptaddress,floatfix]{revtex4}
\usepackage{graphicx} 
\usepackage{amsmath, bbm,amssymb}   
\pdfoutput=1
\begin{document}
\bibliographystyle{prsty}
\title{Dynamical tunneling in macroscopic systems}
\author{I. Serban}  
\affiliation{Department Physik, Arnold-Sommerfeld-Center for Theoretical Physics, and Center for Nanoscience, Ludwig-Maximilians-Universit\"at, Theresienstr.~37, 80333 M\"unchen, Germany} 
\affiliation{IQC and Dept.~of Physics and Astronomy, University of Waterloo, 200 University Ave W, Waterloo, ON, N2L 3G1, Canada}  
\author{F.~K.~Wilhelm}  
\affiliation{IQC and Dept.~of Physics and Astronomy, University of Waterloo, 200 University Ave W, Waterloo, ON, N2L 3G1, Canada}  
\date{\today}
\begin{abstract}  
We investigate macroscopic dynamical quantum tunneling (MDQT) in the
driven Duffing oscillator, charateristic for Josephson junction physics
and nanomechanics. Under resonant conditions between stable coexisting states of such systems we calculate the tunneling rate. In macroscopic systems coupled to a heat bath, MDQT can be masked by driving-induced activation. We compare both processes, identify conditions under which tunneling can be detected with present day experimental means and suggest a protocol for its observation. 
\end{abstract}
\maketitle
The phase space of a classical system can have forbidden areas even in the
absence of potential barriers, e.g.~in the presence of external driving. Quantum-mechanically, these areas can be crossed in a process called 
dynamical tunneling \cite{Heller99, Heller81}. So far, dynamical tunneling has been observed experimentally in microscopic systems, i.e.~cold atoms \cite{Hensinger01Steck01} with very low dampig. Recent experimental
progress has demonstrated many basic quantum features in macroscopic systems
such as Josephson junctions or nanomechanical oscillators,
overcoming the limitations posed by their coupling to the environment. Important for this success was the ability to reduce noise and cool to very low temperatures.  

In this paper we discuss the possibility of macroscopic dynamical tunneling (MDQT) i.e.~involving a macroscopic degree of freedom, like the phase difference across a driven Josephson junction. Classically, for certain parameters, this system has two stable coexisting oscillations with different amplitudes.
This driven system will
feel the influence of its dissipative environment strongly even at temperature
$T=0$. We demonstrate that under experimentally accessible conditions the tunneling between the two classical states can indeed occur and be singled out from the background of classical switching events. We suggest an experiment where MDQT can be directly observed. Our result can be applied  to verify quantum physics in systems with weak nonlinearity such as nanomechanical oscillators. Quantum tunneling it is also a potential dark count error process in the Josephson bifurcation amplifier. Here the classical switching between the two driving-induced, coexisting states in a Josephson junction was used for high resolution dispersive qubit state detection 
\cite{Siddiqi04, Lupascu06, Lee07, Siddiqi06b}. 

Dynamical tunneling (in the absence of an environment) has been studied using WKB in the parametric driven oscillator \cite{Dykman07}. Activation rates in the presence of an environment have been studied in bistable systems \cite{Dykman06, Dykman06b88}. Dynamical tunneling with dissipation has been described numericaly \cite{Peano06b} and multiphoton resonances have been studied perturbatively \cite{Peano06}.

We study a harmonically driven Duffing oscillator, as an approximate description of a wide range of macroscopic physical systems ranging Josephson junctions \cite{Siddiqi05, Siddiqi04} and nanomechanical oscillators \cite{Almog07, Aldridge05}.  
The driven Duffing oscillator is described by the Hamiltonian
\begin{eqnarray}
\hat{H}(t)&=&\frac{\hat{p}^2}{2m}+\frac{m\Omega^2}{2}\hat{x}^2-\gamma \hat{x}^4+F(t)\hat{x},\label{ham1}
\end{eqnarray}
where $F(t)=F_0(\mathbbm{e}^{\mathbbm{i}\nu t}+\mathbbm{e}^{-\mathbbm{i}\nu t})$ is the driving field with frequency $\nu$. For sub-resonant driving $\nu<\Omega$ and below a critical driving strength $F_0<F_c$ two classical oscillatory states with different response amplitudes coexist. Considering a Josephson junction with capacitance $C$, critical current $I_c$ and driving current amplitude $I$ we can identify $x$ as the
phase difference across the junction,
$m=(\hbar/2 e)^2C$, $\Omega=\sqrt{2 e I_c/(\hbar C)}$, $F_0=\hbar I/(2e)$ and $\gamma=m\Omega^2/24$. 

Following the Caldeira-Leggett approach, we assume an Ohmic environment and describe it as a bath of harmonic oscillators 
\begin{equation*}
\hat{H}_E=\sum_i\left(\frac{m_i\omega_i^2\hat{x}_i^2}{2}+\frac{\hat{p}_i^2}{2m_i}\right)-\hat{x}\sum_i\lambda_i\hat{x}_i+\hat{x}^2\sum_i\frac{\lambda_i^2}{2m_i\omega_i^2},
\end{equation*}
with spectral density $J(\omega)=\pi\sum_i\lambda_i^2\delta(\omega-\omega_i)/(2m_i\omega_i)=m\kappa\omega \exp(-\omega/\omega_c)$ and $\omega_c$  a high frequency cutoff.

We transform this Hamiltonian using the the unitary operator $\hat{U}=\exp(\mathbbm{i}\nu t(\hat{a}^{\dagger}\hat{a}+\sum_i\hat{b}_i^{\dagger}\hat{b}_i)$ similar to Ref.~\cite{Dykman06}, where $\hat{a}$ and $\hat{b}_i$ are the annihilation operators for the system and bath oscillators. Dropping the fast rotating terms in the rotating wave approximation (RWA), we obtain 
\begin{equation}
\hat{H}_{\rm tot}=\hat{H}_0^{(\delta)}-x\sum_i\lambda_i\hat{x}_i+\sum_i\frac{\tilde{m}\tilde{\omega}_i^2\hat{x}_i^2}{2}+\frac{\hat{p}_i^2}{2\tilde{m}_i},\label{hamtot}
\end{equation}
where, up to a constant we have
\begin{equation}
\hat{H}_0^{(\delta)}=\frac{\tilde{m}\tilde{\Omega}^2}{2}\hat{x}^2+\frac{\hat{p}^2}{2\tilde{m}}-\frac{6\gamma}{4\tilde{m}^2\tilde{\Omega}^4}\left(\frac{\tilde{m}\tilde{\Omega}^2}{2}\hat{x}^2+\frac{\hat{p}^2}{2\tilde{m}}\right)^2+F_0\hat{x}\label{ham0}.
\end{equation}
We thus obtain a time independent Hamiltonian at the expense of a form that is not separable in $\hat{p}$ and $\hat{x}$. This transformation reduces the frequency $\tilde{\Omega}=\Omega\delta$ and increases the mass $\tilde{m}=m/\delta$ of the oscillators by $\delta_i=(\omega_i-\nu)/\omega_i$ in the case of the bath and $\delta=(\Omega-\nu)/\Omega+\kappa\omega_c/(\pi\Omega^2)$ for the main oscillator, where the
additional term describes a deterministic force induced by dragging the 
system through its environment.

\begin{figure}[!ht]
\includegraphics[width=.45\textwidth,height=5cm]{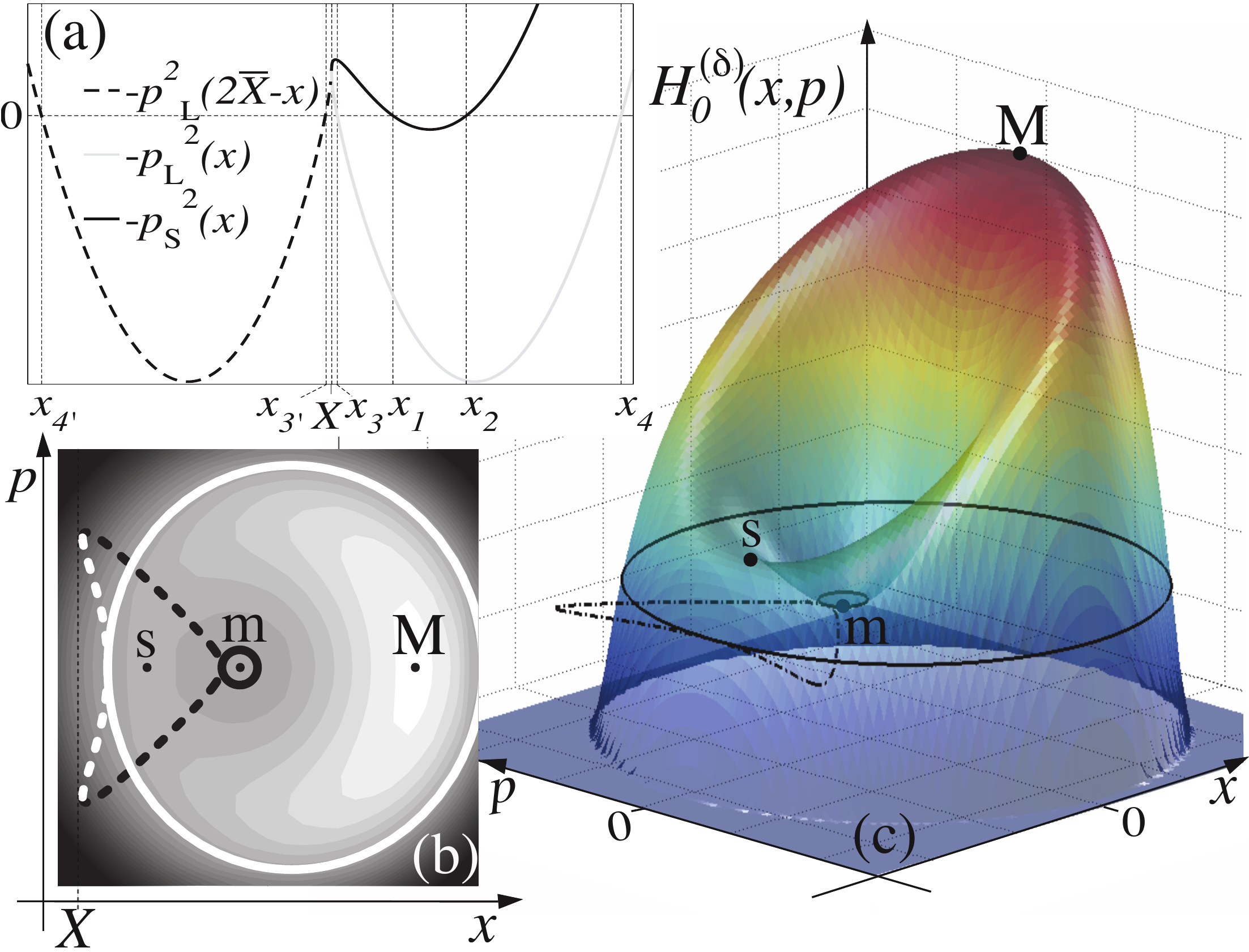}
   \caption{ Illustration of the Hamilton function and the "potential" landscape. (a) $-p_{S,L}^2(x,E)$: the "potential" changes with $E$; classical turning points are found at $p(x_i,E)=0$. (b,c) $H_0^{(\delta)}(x,p)$; in (b) white corresponds to high, black to low quasi-energy; (b): white lines corresponds to the $L$, black ones to the $S$ branch; continuous lines correspond to real and dashed ones to imaginary valued momentum.} \label{cake}
\end{figure}

We concentrate at first on quantum tunneling in the absence of bath fluctuations and study the system in the phase space. The classical Hamilton function $H_0^{(\delta)}(x,p)$ is portraited in
Fig.~\ref{cake}(b) and (c) for a sub-critical driving strength $F_0<F_c=2/9(2\tilde{m}^3 \tilde{\Omega}^6/\gamma)^{1/2}$. It
has three extremal points: saddle (s), minimum (m) and maximum (M) with coordinates $(x_{\rm e},p_{\rm e})$ in the phase space, where ${\rm e}\in\{{\rm m,s,M}\}$. 
The curves satisfying $H_0^{(\delta)}(x,p)=E$ represent classical trajectories. In the following we call $E$ the quasi-energy. In the bistability region $E\in(E_{\rm m},E_{\rm s})$ where $E_{\rm e}=H_0^{(\delta)}(x_{\rm e},p_{\rm e})$ there are always two periodic classical trajectories, around the two stable points (m) and (M), with a small and large amplitude respectively.   

Using this phase space, we outline an experiment to observe MDQT during the transient evolution of the system. Without driving, the system relaxes to its ground
state centered around (m). Then, after switching on the driving field 
one records the time needed for a transition to the large orbit as a function of a driving parameter e.g.~frequency $\nu$.   
When two quantized levels pertaining to the two oscillatory states have almost the same quasi-energy, tunneling can occur, and enhance the total switching rate. 

We describe tunneling using the semiclassical WKB approximation which is an expansion in $\hbar$ close to the least action path. To find that path 
we solve the equation $H_0^{(\delta)}(x,p)=E$ and obtain four coexisting momentum branches $\pm p_{L,S}(x,E)$ where
\begin{eqnarray}
p_{S,L}(x,E)\!\!\!&=&\!\!\!\tilde{m}\tilde{\Omega}\sqrt{\frac{2\tilde{m}\tilde{\Omega}^2}{3\gamma}-x^2\mp \sqrt{\frac{8F_0}{3\gamma}} \sqrt{x-X }},
\end{eqnarray}
with $X ={E}/{F_0}-(\tilde{m}\tilde{\Omega}^2)^2/(6F_0\gamma)$. This configuration is reminiscent of Born-Oppenheimer surfaces in the molecular physics where dynamical tunneling has also been studied \cite{Heller99}. A real-valued
$p_{S,L}$ corresponds to a classically allowed area with an oscillating WKB
wave function, a complex-valued one to a classically forbidden area with
a decaying wave function. 
  At $x=X $, both trajectories have the same momentum and position and connect. Here $\dot x=\partial_pH_0^{(\delta)}(x,p)=0$ but $p\not=0$ such that the the motion changes direction and continues on a different momentum branch.  
For all $x<X $ both $p_{S,L}(x,E)$ are complex, thus this forbidden area does not influence the quantization rules within the WKB approximation. 
The tunneling least-action trajectory which connects the two allowed regions only passes through the region $x>X $. Here the $p_{S,L}$ are either real or purely imaginary, i.e.~$-p_{S,L}^2\in \mathbbm{R}$. To study this region, we mirror the solution $p_{L}(x,E)$ around the $X $ point as shown in Fig.~\ref{cake}(a) and obtain a double well ``potential". The small and large amplitude oscillation states  are localized in the right and left-hand wells, respectively, and are separated by a ``potential barrier" with purely imaginary momentum. Here we do not observe any interference effects, as opposed to the case of the parametrically driven oscillator \cite{Dykman07}.
We apply the WKB theory in this ``potential" in order to determine the tunnel splitting in the limit of a low transmission through the forbidden region. The classical turning points $x_i$ are given by $p_{S,L}(x_i,E)=0$, see Fig.~\ref{cake} (a).  The bound state energies at zero transmission are given by the Sommerfeld energy quantization rules
\begin{equation}
S_{12}(E)=\pi n+\pi/2,\:\:\:\:S_{4'3'}(E)=\pi m+\pi/2,\:\:\:\: n,m\in\mathbbm{Z}, \label{condnotunn}
\end{equation}
where $S_{ij}(E)=\int_{x_i}^{x_j}{\rm sign}(x-X )|p(x,E)|dx/\hbar$ and the negative sign on the left hand side of $X $is due to mirroring. Whenever a pair of energies from either well is degenerate, resonant tunneling through the barrier can occur. This induces coupling between the two wells and lifts the degeneracy. The level crossings become avoided crossings at finite transmission and the full WKB condition reads
\begin{eqnarray}
\cot S_{12}(E)\cot S_{4'3'}(E)=\exp(-2S_{3'1}(E))/4\label{condtunn}.
\end{eqnarray}
We expand the quasi-energy $E$ and the actions $S_{ij}$ in series of 
$\xi=1/4\exp\left(-2S_{3'1}\right)$ around the level crossings with quasi-energy $E_0$ where eqs.~(\ref{condnotunn}) are simultaneously satisfied. 
The first energy correction $E_1 \xi$ is obtained straightforwardly from
$\partial_{E}S_{12}|_{E_0}\partial_{E}S_{4'3'}|_{E_0}(E_1\xi)^2=\xi,$
and the  tunneling rate is obtained directly from the energy splitting at the avoided level crossings
\begin{eqnarray}
\Gamma_{\rm t}&=&\frac{2E_1\xi}{\hbar\pi}=\frac{\exp(-S_{3'1})}{\hbar\pi\sqrt{\partial_{E}S_{12}\partial_{E}S_{4'3'}}}\Bigg|_{E_0}.\label{tunnel}
\end{eqnarray}
This can be evaluated in closed form involving elliptic integrals for $S_{ij}$ and we obtain the exact expressions
 \begin{eqnarray*}
 \partial_{E}S_{12}|_{\rm m}&=&\partial_{E}S_{4'3'}|_{\rm m}=\pi/(\hbar\Omega_{\rm m}),
 \\\partial_ES_{12}|_{\rm s}&=&\partial_ES_{3'1}|_{\rm m}=\infty,\:\partial_{E}S_{3'1}|_{\rm s}=\pi/(\hbar|\Omega_{\rm s}|),
 \end{eqnarray*}
where $\Omega_{\rm e}=\sqrt{\partial^2_{xx}H_0^{(\delta)}\partial^2_{pp}H_0^{(\delta)}}|_{\rm e}$ and ${\rm e}\in\{{\rm m,s,M}\}$. Thus, for $S_{12}$ at (m) and $S_{3'1}$ at (s) we reproduce the harmonic oscillator result. The saddle point ``frequency" $\Omega_{\rm s}$ is imaginary as expected. 

We simplify Eq.~(\ref{tunnel}) by locally approximating $H_0^{(\delta)}$ close to the extremal points by harmonic oscillators, i.e.~assuming that $S_{ij}$ are linear functions of $E$. This approximation holds for all $S_{ij}$ simultaneously when $E$ is far enough from both extremal points $E_{\rm s,m}$, as it is the case for the ground state $E_{\rm m}+\hbar \Omega_{\rm m}/2$ of the small amplitude well. In this approximation $S_{3'1}(E)\approx\pi(E_{\rm s}-E)/(\hbar|\Omega_{\rm s}|)$ and thus we find a compact approximation
\begin{equation}
\Gamma_{\rm t}\approx\Omega_{\rm m}/\pi^2\exp(-\pi(E_{\rm s}-E_{\rm m}-\hbar \Omega_{\rm m}/2)/(\hbar|\Omega_{\rm s}|)).\label{approx_actions}
\end{equation}
\begin{figure}
  \includegraphics[width=0.45\textwidth,height=5cm]{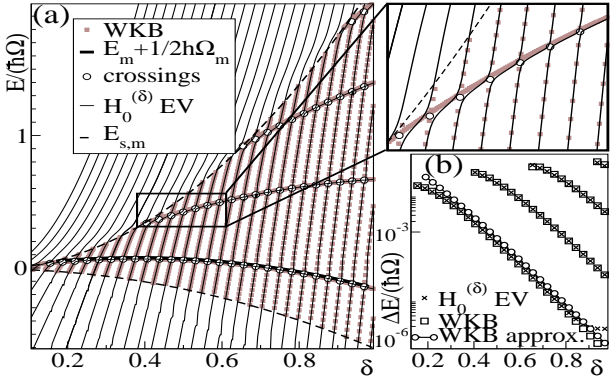}
     \caption{(a): Quantized energies: eigenvalues (EV) of $\hat{H}_0^{(\delta)}$ versus WKB. $\hat{H}_0^{(\delta)}$ was represented in the number state basis considering a number of levels $2N$. (b): Tunneling-induced energy splittings at level crossings. Frequency sweep at $m\Omega/\hbar=2$, $\gamma=m\Omega^2/24$, 
$\kappa\omega_c/\Omega^2=0.1$ and $F_0=0.5 F_c(\nu)$.}\label{energies}
 \end{figure}
Our calculations rely on a series of assumptions. To test them, we compare the
results to a full numerical diagonalization of $\hat{H}_0^{(\delta)}$ 
taking a basis of the first $2N$ Fock states. At $F_0=0$, the number of levels that cover the bistability region is $N=\hbar\Omega(2m\Omega)^2/(6\gamma\hbar^2)$. As shown for a representative
set of data in  Fig.~\ref{energies}, we find good agreement between these
numerically exact results and the predictions of Eqs.~(\ref{condnotunn},\ref{tunnel}) and
also (\ref{approx_actions}).

Quantum tunneling is significant only close to level 
crossings. It always competes with the activation over the barrier, which 
occurs at all energies and is based on classical fluctuations due
to coupling to a heat bath. A rather detailed treatment of a similar process has been given in Refs.~\cite{Dykman06b88}. We now estimate these effects and compare them to 
the quantum tunneling rate. When modeling activation, it is crucial to consider that we are working in a frame rotating relative
to the heat bath, which is fixed in the laboratory. 

We start from Eq.~(\ref{hamtot}). As we will adopt the mean first passage
time approach \cite{Hanggi90}, it is sufficient to approximate the system Hamiltonian close to its minimum in phase space by 
$\hat{H}_0^{(\delta)}\approx\hat{p}^2/(2 m_{\rm eff})+V(\hat{x})$ where the effective mass is determined by the curvature of the Hamilton function $m_{\rm eff}^{-1}=\partial_{pp}^2H_0^{(\delta)}(x,p)|_{\rm m}$ and the effective potential is $V(x)=H_0^{(\delta)}(x,p_{\rm m})$. 
In this approximation we obtain a quantum Langevin equation
\begin{equation}
m_{\rm eff}\ddot{x}+\partial_xV(x)-x\!\!\!\int_0^{\infty}\!\!\!\!\!d\omega\frac{2J(\omega)}{\pi\omega}+m_{\rm eff}\!\!\int_0^t\!\!\tilde\kappa(t-s)\dot{x}(s)ds=\xi(t),\nonumber
\end{equation}
where 
\begin{eqnarray*}
\tilde\kappa(t)&=&\int_0^{\infty}\frac{2 J(\omega)\cos((\omega-\nu)t)}{(\omega-\nu)\pi m_{\rm eff}}d\omega,\\
\xi(t)&=&\sum_i\lambda_i\left[\!\!\left(x_i(0)-\frac{\lambda_i x(0)}{\tilde m_i\tilde \omega_i^2}\right)\cos(\tilde{\omega}_i t)+\frac{p_i(0)}{\tilde{m}_i\tilde{\omega}_i}\sin(\tilde{\omega}_i t)\!\right]\!.
\end{eqnarray*}
$\tilde{\kappa}(t)$ is peaked on a short time scale $\omega_c^{-1}$. Its magnitude is characterized through the effective friction constant 
\begin{equation*}
\kappa_{\rm eff}=\int_0^{\infty}\!\!\!\!\!\!\tilde\kappa(t)dt=2\kappa\left(\delta-\frac{3\gamma x_{\rm m}^2}{2 m\Omega^2}\right)(1+\mathcal{O}(\nu/\omega_c)).
\end{equation*}
The factor of two difference between $\kappa_{\rm eff}$ and the damping constant of the
undriven harmonic system accounts for the fact that in the rotating frame there are bath modes above and below $\omega=0$ (see Eq.~(\ref{hamtot})) whereas for the undriven case the frequencies are strictly positive.  
Thus oscillators with frequency $\omega$ have the spectral density $J(\omega+\nu)$ and modes with negative frequencies have significant contribution to noise even at low temperatures. We use a detailed balance condition to determine the effective temperature of the bath as seen by a detector in the rotating frame, e.g.~a two level system with level separation $\hbar\Omega_{\rm m}$ 
\begin{eqnarray}
P(\Omega_{\rm m},T)/P(-\Omega_{\rm m},T)=\exp\left(\hbar\Omega_{\rm m}\beta_{\rm eff}\right).
\end{eqnarray}
Here $P(\omega,T)=J(\omega+\nu)(1+n(\omega+\nu,T))$ is the probability for a quantum $\hbar\omega$ to be emitted to the bath in the rotating frame. The increase of the quasi-energy in the rotating frame can actually lower the energy in the laboratory frame, i.e.~what a detector in the rotating frame regards as absorption can actually be emission in the lab frame. In the case of constant acceleration in relativistic context this behavior is known as the Unruh effect \cite{Unruh76}.

\begin{figure}[!ht]
 \hspace{-0.45cm} \includegraphics[width=0.25\textwidth]{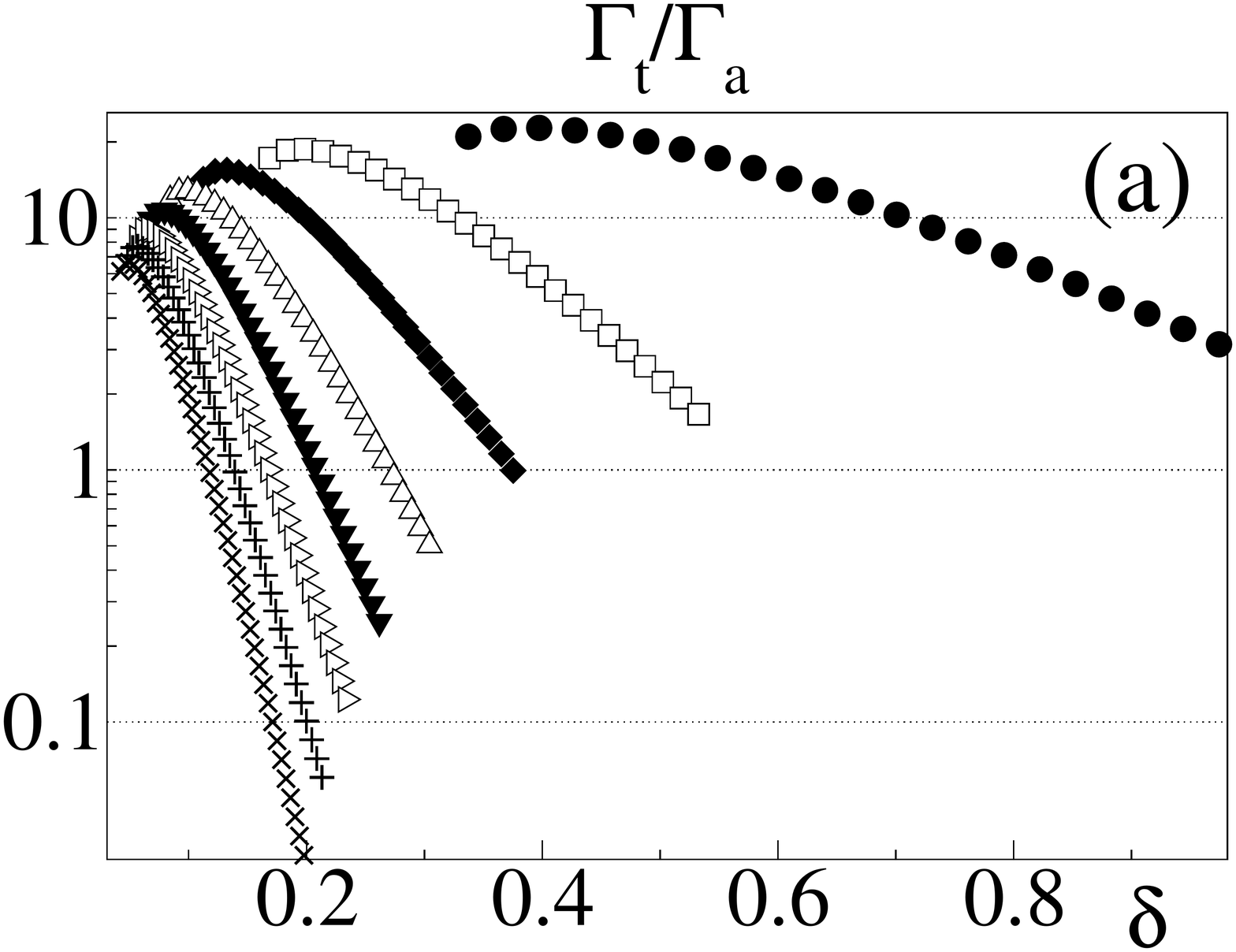} 
 \hspace{-0.25cm}    \includegraphics[width=0.25\textwidth]{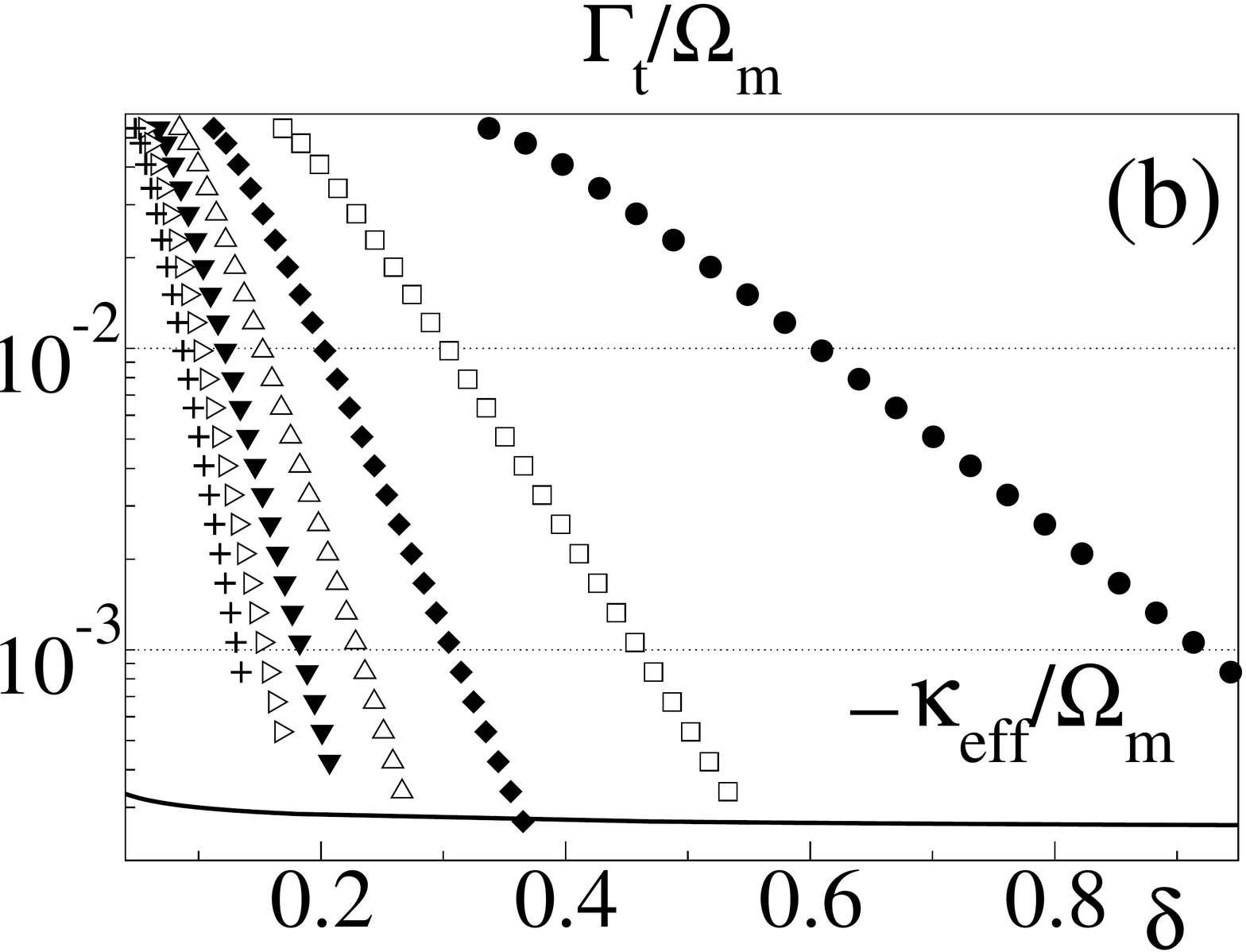}
    \caption{(a) The ratio of tunneling and activation rates from the small well at the avoided level crossings. (b) Corresponding tunneling rates compared to $\kappa_{\rm eff}$ (where $\Gamma_{\rm t}/\Gamma_{\rm a}>1$). Driving frequency sweep at $F_0=0.7F_c(\nu)$; Values of $\Omega$ (in GHz): 1($\bullet$), 2({\tiny$\square$}), 3({\tiny$\blacklozenge$}), 4({\tiny$\triangle$}), 5({\footnotesize$\blacktriangledown$}), 6($\triangleright$), 7(+), 8($\times$), at parameters specified in text.
}\label{rate}
\end{figure}

The barrier crossing problem for systems described by a quantum Langevin equation is well studied in the context of chemical reactions. For low damping,
$\kappa_{\rm eff}\ll \Omega_{\rm m}$ mean-first-passage time theory predics the activation rate \cite{Hanggi90} 
\begin{equation}
\Gamma_{\rm a}^{-1}=\frac{\beta_{\rm eff}}{\kappa_{\rm eff}}\int_0^{S(E_{\rm s})}d S\mathbbm{e}^{-\beta_{\rm eff}E(S)}\int_{E(S)} ^{E_{\rm s}}d E'\frac{\mathbbm{e}^{-\beta_{\rm eff}E'}}{S(E')}, \label{integral}
\end{equation}
where $S(E)=\oint p(x,E) dx$.  
In the traditional low temperature limit $\kappa_{\rm eff} S(E_{\rm s})\ll k_B T_{\rm eff}\ll E_{\rm s}-E_{\rm m}$ the activation rate becomes 
\begin{equation}
\Gamma_{\rm a}=\kappa_{\rm eff}\beta_{\rm eff}\Omega_{\rm m}\exp\left(-(E_{\rm s}-E_{\rm m})\beta_{\rm eff}\right)S(E_{\rm s})/(2\pi).
\end{equation}
In our case, the noise temperature $k_BT_{\rm eff}$ can be larger than the barrier height $E_{\rm s}-E_{\rm m}$. In this limit we obtain from Eq.~(\ref{integral})
\begin{equation}
\Gamma_{\rm a}=\kappa_{\rm eff}(F(\beta_{\rm eff}(E_{\rm s}-E_{\rm m})))^{-1}\label{activation}
\end{equation}
where $F(x)=\int dx (\exp(x)-1)/x\equiv{\rm Ei}(x)-\log(x)$. 

Summarizing, in the rotating frame, as a consequence of driving, the bath appears with a quality factor $\Omega_{\rm m}/\kappa_{\rm eff}$ reduced by approximatively a factor of two and an enhanced effective temperature $T_{\rm eff}$. Moreover, the bath shifts the detuning $\delta$. We show that experimental observation of MDQT could still be possible.
At the the level anticrossings we calculate the WKB tunneling rate from the ground state and the activation rate from Eq.~(\ref{activation}), see Fig.~\ref{rate}(a) where we have considered a Josephson junction with $\kappa=10^{-4}\Omega$, the temperature $T=10$ mK, shunt capacitance $C=2\cdot10^{-12}$~F and $\gamma=m\Omega^2/24$. The values of $\delta$ where these anticrossings occur are found by minimizing $|\cot(S_{4'3'}(E_{\rm m}))|$ and are in agreement with the weak driving result \cite{Peano06} $\delta= 3\gamma n/(2 m^2\Omega^3)$, $n\in\mathbbm{N}$. 
We observe that the quantum tunneling rate can be one order of magnitude larger than the activation rate in the limit of relatively small detuning $\delta$ and low damping. By increasing the value of $\alpha=m\Omega/\hbar$, we observe a reduction of the ratio $\Gamma_{\rm t}/\Gamma_{\rm a}$ as expected, since $\alpha $ measures the number of quantized levels in the system and thus the "classicality" of its behavior. In Fig.~\ref{rate} we have $\alpha\in(2,20)$, while in the experiment of Ref.~\cite{Siddiqi05} 
$\alpha$ was larger than 100, at higher temperature and smaller quality factor, such that MDQT was probably masked by thermal activation. We expect that at the values of Fig.~\ref{rate} the experiment we propose should produce direct evidence for MDQT.

In conclusion we have investigated macroscopic dynamical tunneling by mapping it onto tunneling between two potential surfaces and found a compact analytic expression for the tunneling rate. We compared this process with the activation over the barrier using the mean first passage time approach. The values obtained suggest that dynamical tunneling can be singled out from the background of activation processes. We have proposed an experiment realizable within existing technology to demonstrate dynamical tunneling by monitoring the switching rate between the two dynamical states while tuning a parameter of the external driving. 

We are thankful to A.~Leggett for pointing out the Unruh effect analogy and to M. Dykman, M. Marthaler, and E.M. Abdel-Rahman for useful remarks. This work was supported by DFG through SFB 631, by 
NSERC discovery grants, and by EU through EuroSQIP. 

\end{document}